\documentclass[preprint,12pt]{elsarticle}

\usepackage{amssymb}
 \usepackage{amsthm}

\usepackage{xspace}
\usepackage{color}
\usepackage{msc}
\usepackage{array}
\usepackage{multirow}
\usepackage{centernot}
\usepackage{mathtools}
\usepackage{stmaryrd}
\usepackage[utf8]{inputenc}

\definecolor{gris}{gray}{0.85}

\newtheorem{example}{Example}
\newtheorem{definition}{Definition}

\makeatletter
\newcommand{\rightdoublearrow}{%
  \rightarrow\mkern-10mu\protect\joinrel\rightarrow}

\newcommand{\rightdoublearrowfill@}
  {\arrowfill@\relbar\relbar\rightdoublearrow}
\newcommand{\mapstodoublearrowfill@}
  {\arrowfill@{\mapstochar\relbar}\relbar\rightdoublearrow}
\newcommand{\xrightdoublearrow}[2][]
  {\ext@arrow 3{15}59\rightdoublearrowfill@{#1}{#2}}
\newcommand{\xmapstodoublearrow}[2][]
  {\ext@arrow 3{15}59\mapstodoublearrowfill@{#1}{#2}}
\makeatother

\newcommand{\SPEC}{\textsc{Spec}}
\newcommand{\APTE}{\textsc{Apte}}
\newcommand{\AKISS}{\textsc{Akiss}}
\newcommand{\tamarin}{\textsc{Tamarin}}
\newcommand{\maudenpa}{\textsc{Maude-NPA}}
\newcommand{\proverif}{\textsc{ProVerif}}

\newcommand{\LRstep}[1]{{\xRightarrow{#1}}}

\newcommand{\ie}{\textit{i.e.,} }
\newcommand{\eg}{\textit{e.g.,} }
\newcommand{\etc}{\textit{etc.}}

\makeatletter
\def\rightarrowfillstar@{\arrowfill@\relbar\relbar{\rightarrow\smash{^*}}}
\newcommand{\xrightarrowstar}[2][]{\ext@arrow
  0{13}{15}8\rightarrowfillstar@{#1}{#2}}
\newcommand{\lrstep}{\@ifstar{\xrightarrowstar}{\xrightarrow}}
\makeatother

\newcommand{\choice}{\mathsf{choice}}
\newcommand{\fst}{\mathsf{fst}}
\newcommand{\snd}{\mathsf{snd}}

\newcommand{\proj}{\mathsf{proj}}

\newcommand{\new}{\mathsf{new}}
\newcommand{\Out}{\mathsf{out}}
\newcommand{\In}{\mathsf{in}}

\newcommand{\sign}{\mathsf{sign}}

\newcommand{\defi}{\mathsf{def}}

\newcommand{\h}{\mathsf{h}}
\newcommand{\vk}{\mathsf{vk}}
\newcommand{\pk}{\mathsf{pk}}

\newcommand{\senc}{\mathsf{senc}}
\newcommand{\mac}{\mathsf{mac}}
\newcommand{\sdec}{\mathsf{sdec}}

\newcommand{\DG}{\mathit{dg}}
\newcommand{\SOD}{\mathit{sod}}
\newcommand{\KPrAA}{\mathit{sk_P}}
\newcommand{\KPrDS}{\mathit{sk_{DS}}}

\newcommand{\KE}{\mathit{ke}}
\newcommand{\KM}{\mathit{km}}
\newcommand{\getC}{\mathsf{get\_Challenge}}
\newcommand{\New}{\mathsf{new}}
\newcommand{\rfstar}{\mathsf{Rf^*}}

\newcommand{\N}{\mathcal{N}}
\newcommand{\X}{\mathcal{X}}
\newcommand{\W}{\mathcal{W}}
\newcommand{\T}{\mathcal{T}}
\newcommand{\vars}{\mathit{vars}}
\newcommand{\dom}{\mathit{dom}}
\newcommand{\E}{\mathsf{E}}

\newcommand{\pub}{\mathsf{pub}}
\newcommand{\priv}{\mathsf{priv}}
\newcommand{\unblind}{\mathsf{unblind}}
\newcommand{\blind}{\mathsf{blind}}
\newcommand{\checksign}{\mathsf{checksign}}

\newcommand{\p}{\mathcal{P}}

\newcommand{\tr}{\mathsf{tr}}
\newcommand{\trace}{\mathsf{traces}}

\newcommand{\q}{\mathcal{Q}}

\newcommand{\config}[2]{(#1 ; #2)}
\newcommand{\myIf}{\mathsf{if }}
\newcommand{\myThen}{\mathsf{then }}
\newcommand{\myElse}{\mathsf{else }}

\begin{document}

\begin{frontmatter}



\title{A survey of symbolic methods for establishing equivalence-based properties in
  cryptographic protocols}


\author[irisa]{St\'ephanie Delaune}
\author[lsv]{Lucca Hirschi}
\address[irisa]{CNRS/IRISA, Rennes, France}
\address[lsv]{LSV, CNRS \& ENS Cachan, France}
\begin{abstract}
Cryptographic protocols aim at securing communications over insecure
networks such as the Internet, where dishonest users may listen to
communications and interfere with them. A secure communication has a
different meaning depending on the underlying application. It ranges
from the confidentiality of a data 
to \emph{e.g.} verifiability in electronic voting systems. Another example of
a security notion is \emph{privacy}.

Formal symbolic models have proved their usefulness for analysing the security
of protocols. 
Until quite recently, most results focused on trace properties like confidentiality or
authentication. There are however several security properties, which
cannot be defined (or cannot be naturally defined) as trace properties
and require a notion of behavioural equivalence. 
Typical examples are anonymity, and privacy related properties. During
the last decade, several results and verification tools have been
developed to analyse equivalence-based security properties.

We propose here a synthesis of decidability and undecidability results
for equivalence-based security properties.
Moreover, we give an overview  of existing verification tools that may
be used to verify equivalence-based security properties.
\end{abstract}


\end{frontmatter}

\section{Introduction}
\label{sec:intro}


Security protocols are widely used to secure transmissions in various types of networks
(\eg web, wireless devices, \etc). They are (often small) concurrent programs relying on
cryptographic primitives. The security properties they should achieve are multiple
and depend on the context in which they are used.
The main problem they have to cope with is to protect communication
that are done through insecure, public, channels
like the Internet, where dishonest users may listen to communications and interfere with them.
This explains why they are notoriously difficult to design and hard to analyse by hand.
Actually, many protocols have been shown to be flawed several years after their publication (and
deployment). Given the very sensitive contexts in which they are used, 
establishing the security of
these protocols is a very relevant research goal with important
economic and societal consequences.

\medskip{}

Two main distinct approaches have emerged, starting with the early
1980's attempt of~\cite{dolev81security},
to ground security analysis of protocols on firm,
rigorous mathematical foundations. These two approaches are known as
the {\em computational approach} and the {\em symbolic approach}. 

The {computational approach} models messages as bit-strings;
agents and the attacker as probabilistic polynomial time machines; whereas
security properties are defined using games played by the attacker who
has to be able to distinguish the protocol from an idealised version
of it (with a non negligible probability).
It is generally acknowledged that security proofs in this
model offer powerful security guarantees. 
A serious downside of this approach however is that even for small
protocols, proofs
are usually long, difficult, tedious, and highly error prone.
Moreover, due to the high complexity of such a model, automating such proofs is a very complex problem
that is still in its infancy (see \emph{e.g.}~\cite{DBLP:conf/sp/Blanchet06}).

By contrast, the symbolic approach, which is the one targeted by this
survey, makes strong assumptions on cryptographic primitives (\ie black-boxed cryptography assumption)
but fully models agents' interactions and algebraic properties of
these primitives.
For instance, symmetric encryption and decryption are modelled as 
function symbols $\mathsf{enc}$ and $\mathsf{dec}$ along with the
equations 
$\mathsf{dec}(\mathsf{enc}(m,k),k) = m$. This means that, 
without the corresponding key~$k$, it is simply impossible to get
back the plaintext~$m$ from the cipher-text $\mathsf{enc}(m,k)$.
This does not mean however that
protocols relying on these primitives are necessarily secure.
There can still remain some {\em logical attacks} like \emph{e.g.} a
man-in-the-middle  attack or a reflection attack.
Although less precise, this symbolic approach 
benefits from automation and can thus 
target more complex protocols than those analysed
using the computational approach.
Moreover, a line of work known
as {\em computational soundness} aims at spanning the gap between these
two approaches by establishing that, in some cases, 
security guarantees in the symbolic model imply security guarantees in the
computational one. This line has been initiated by
Abadi and Rogaway~\cite{DBLP:conf/ifipTCS/AbadiR00} and has received much attention since then
(see~\cite{cortier2011survey} for a survey on computational soundness).

\medskip{}


Until the early 2000s, most works from the symbolic approach were focusing
on {\em trace properties}, that is, statements that something bad never
occurs on any execution trace of a protocol. Secrecy and
authentication are typical examples of trace properties: a data
remains confidential if, for any execution, the attacker is
not able to produce the data from its observations.
But many other properties like
strong secrecy, unlinkability or anonymity are not defined as trace
properties. These properties are usually defined as the fact that
an observer cannot distinguish between two situations, and
require a notion of {\em behavioural equivalence}. Roughly, two
protocols are equivalent if an attacker cannot observe whether he is
interacting with 
one or the other.
In this survey, we shall focus on equivalence-based security properties.


There exist other approaches out of the scope of this survey
that do not strictly follow the symbolic approach nor the
computational one but are able to verify notions of behavioural
equivalence.
A recent approach proposes to define a computationally complete symbolic attacker
by 
axiomatizing what the attacker can \emph{not} do~\cite{bana2012towards}. 
This approach has been recently extended to deal with a notion of behavioural equivalence~\cite{bana2014computationally}.
Another work proposed semi-automatic proof of
vote privacy using type-based verification~\cite{cortier2015type}. This has been done using the tool $\rfstar$,
where protocols are modelled using code-based cryptographic abstractions and
security properties are encoded as {\em refinement
  types}~\cite{barthe2014probabilistic}. Security is achieved by type checking the protocol.

\paragraph{Outline}
In Section~\ref{sec:protocol}, we give
an informal presentation of different cryptographic primitives after
which we describe the  Basic Access Control (BAC) protocol
from the e-passport application,  and some of its logical attacks.
Section~\ref{sec:properties} describes various security properties
that one may want to verify.
In Section~\ref{sec:model},
we give a formal model, {following the symbolic approach},
for messages, protocols and equivalence properties.
Section~\ref{sec:methods} is dedicated to the existing methods and tools for verifying equivalence-based properties.
We conclude in Section~\ref{sec:conclu}.

\section{What is a cryptographic protocol?}
\label{sec:protocol}
A cryptographic protocol can be seen as a list of rules that describe executions;
these rules specify the emissions and receptions of messages by the
actors of the protocols called \emph{agents}. These protocols
use as basic building blocks cryptographic primitives such as symmetric/asymmetric
encryptions, signatures, and hash functions. For a long time, it was
believed that designing a strong encryption scheme was sufficient to
ensure secure message exchanges. Starting from the 1980's, researchers
understood that even with perfect encryption schemes, message
exchanges were still not necessarily secure. This fact will be
illustrated in Section~\ref{subsec:protocol}, but we first briefly
explain the most standard cryptographic primitives together with their
fundamental properties. 


\subsection{Cryptographic primitives}
\label{subsec:primitives}

Cryptographic primitives provide fundamental properties and are
used to develop more complex tools called cryptographic protocols,
which guarantee one or more high-level security properties.

\paragraph*{Symmetric encryption}
Symmetric cryptography refers to encryption methods in which both
the sender and the receiver share the same key. 
For instance, the Data Encryption Standard (DES) and the Advanced
Encryption Standard (AES) are symmetric encryption schemes 
which have been
designated cryptography standards by the US government in 1976 and
2002 respectively.

A significant disadvantage of symmetric ciphers is the key management
necessary to use them securely. Each distinct pair of communicating
parties must 
share a different key.
Therefore, the number of required keys  increases as
the square of the number of network members, which very quickly
requires complex key management schemes to keep them all straight and
secret. The difficulty of securely establishing a secret key between
two communicating parties, when a secure channel does not already
exist between them, also presents a chicken-and-egg problem which is a
considerable practical obstacle for the use of cryptography in 
the real world. This is why the recourse to  asymmetric
cryptography is so popular for key establishment protocols that aim to
establish a fresh symmetric key between two parties.

\paragraph*{Asymmetric encryption}
In 1976, Diffie and Hellman
proposed the notion of public key 
cryptography, in which two different but mathematically
related keys are used -- a public key and a private key~\cite{DiffieHelman76}. A public key
system is  constructed, in such a way that calculation of one key 
(the 'private
key') is computationally infeasible from the other (the 'public key'),
even though they are necessarily related. 
In public key cryptosystems, the public key may be freely distributed,
while its associated  private key must remain secret. The public key is
typically used for encryption, while the private key is used
for decryption. Diffie and Hellman showed 
that public key cryptography was possible\footnote{Diffie and Hellman have been rewarded
    by the ACM Turing Award in 2015 for having laid the foundations of asymmetric encryption.}
 by presenting the Diffie-Hellman key exchange protocol~\cite{DiffieHelman76}.
In~1978, Rivest, Shamir, and Adleman invented RSA, another
public key cryptosystem~\cite{RSA78} which has established itself as
the main standard.

\paragraph*{Digital signature}
Over the same period, signature schemes have  also been proposed. 
A digital signature is a mathematical scheme for demonstrating the
authenticity of a digital message or of a document. It 
gives the recipient a reason to believe that the message was created by a
known sender, 
that the sender cannot deny having sent the message (authentication
and non-repudiation), 
and that the message was not altered while in transit (integrity). 
Digital signatures are commonly used for software distribution, key management,
financial transactions, \etc

\paragraph*{Hash function}
A hash function takes a message of any length as input, and outputs a
short, fixed length hash. 
Hash functions have many information security applications, notably in digital
signatures, message authentication codes (MACs), and other forms of
authentication. They can also be used as checksums to detect accidental data corruption.
For good hash functions, an attacker cannot find two
messages that produce the same hash. 
Message authentication codes are much like cryptographic hash
functions, except that a secret key can be used to authenticate the
hash value upon receipt.

This list of cryptographic primitives is not exhaustive, and modern protocols often rely on less
standard cryptographic primitives, such as blind
signature, homomorphic encryption, trapdoor bit commitment.

\subsection{An example: the BAC protocol}
\label{subsec:protocol}

For the purpose of illustration, we consider the Basic Access Control (BAC)
protocol used in the
e-passport application. An e-passport is a paper passport with an RFID chip that stores the
critical information 
printed on the passport. The International Civil Aviation Organisation
(ICAO) standard
 specifies the communication protocols that are used to access this information~\cite{ICAO-Passport}.
We do not plan to describe all the protocols that are specified in the
standard. Instead, we shall
concentrate only on  the BAC protocol following the modelling
proposed in~\cite{arapinis-csf10}.

\medskip{}

The information stored in the chip is organised in data groups
($\DG_1$ to~$\DG_{19}$). For example, $\DG_5$ contains a JPEG copy of
the displayed picture, and~$\DG_7$ contains the displayed
signature. The verification key~$\vk(\KPrAA)$ of the passport,
together with its certificate $\sign(\vk(\KPrAA), \KPrDS)$ issued by
the Document Signer Authority, is stored in~$\DG_{15}$. The
corresponding signing key~$\KPrAA$ is stored in a tamper resistant
memory, and cannot be read or copied. 
For authentication purposes, a hash of all the data groups, together
with a signature 
on this hash value issued by the Document Signer Authority, are stored in a separate file, 
the Security Object Document:
\[
\SOD \stackrel{\defi}{=} \langle \sign(\h(\DG_1, \dots, \DG_{19}), \KPrDS), \; \h(\DG_1, \dots, \DG_{19}) \rangle.
\] 
The ICAO standard specifies several protocols through which this
information can be accessed. In particular, read access to the data on the
passport is protected by the BAC protocol.

\begin{figure*}[ht]
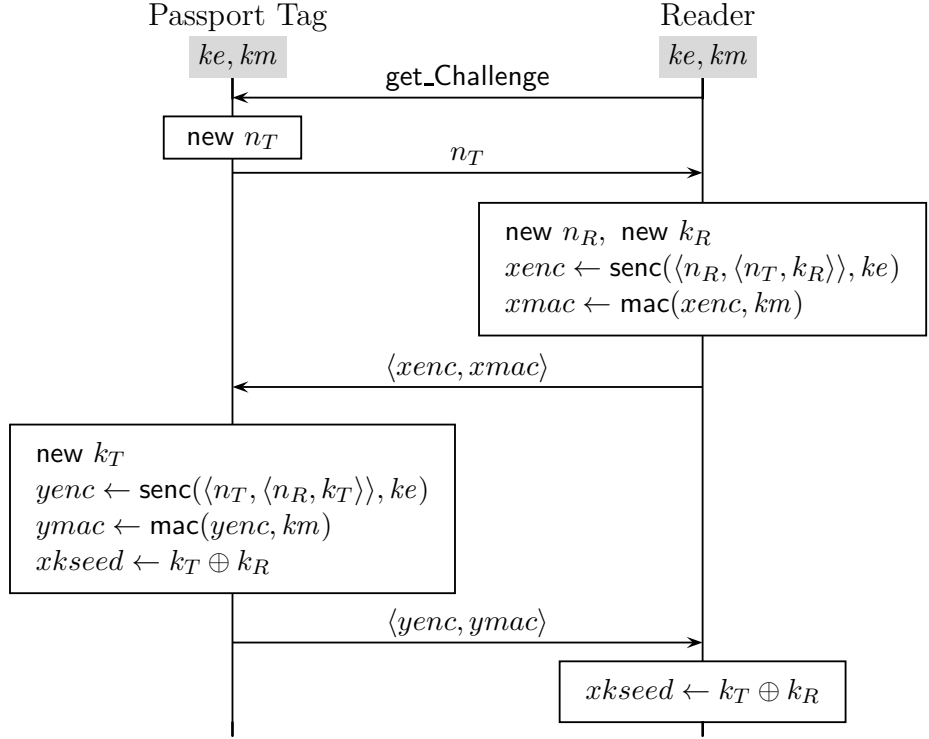

  \centering
  \setmsckeyword{} 
  \drawframe{no}
  \begin{msc}{}
    \setlength{\instwidth}{0\mscunit}
    \setlength{\instdist}{1.25cm} 
    \dummyinst{d0}
    \declinst{pport}{
      \begin{tabular}[c]{c}
        Passport Tag \\
        \colorbox{gris}{\small{$\KE, \KM$}}
      \end{tabular}}{}
    
    \dummyinst{d1} 
    \dummyinst{d2} 
    \dummyinst{d3} 
    \dummyinst{d4}
    
    \declinst{reader}{
      \begin{tabular}[c]{c}
        Reader \\
        \colorbox{gris}{\small{$\KE, \KM$}}
      \end{tabular}}{}
    
    \nextlevel[-1.25]
    \mess{\small{$\getC$}}{reader}{pport}
    \nextlevel[0.5]

    \action*{
      \small{$\New\ n_T$}
    }{pport}
    \nextlevel[1.5]

    \mess{\small{$n_T$}}{pport}{reader}
    \nextlevel[0.8]

    \action*{\small{$
        \begin{array}[c]{l}
          \New\ n_R,\ \New\ k_R \\
          xenc \leftarrow \senc(\langle n_R, \langle n_T, k_R\rangle\rangle, \KE) \\
          xmac \leftarrow \mac(xenc, \KM) \\
        \end{array}
        $}}{reader} 
    \nextlevel[4.9]
    
    \mess{\small{$\langle xenc,  xmac \rangle$}}{reader}{pport} \nextlevel[1]
    
    \action*{\small{$
        \begin{array}[c]{l}
          \New\ k_T \\
           yenc\leftarrow \senc(\langle n_T, \langle n_R, k_T\rangle\rangle, \KE) \\
           ymac\leftarrow \mac(yenc,\KM) \\
           xkseed \leftarrow k_T \oplus k_R \\
        \end{array}
        $}}{pport} 
    \nextlevel[5.8]
    
    \mess{\small{$\langle yenc, ymac\rangle$}}{pport}{reader}
    \nextlevel[0.5]

    \action*{\small{$
        \begin{array}[c]{l}
           xkseed \leftarrow k_T \oplus k_R
        \end{array}
        $}}{reader} 
    \nextlevel[0.8]    
  \end{msc}
  \caption{Basic Access Control protocol
  \label{fig:bac}
}
\end{figure*}

The BAC protocol is a password-based authenticated key exchange protocol (PAKE) whose
security relies on two master keys, namely $\KE$ and $\KM$, which are derived from a password of
low entropy optically retrieved from the passport by the reader before executing the protocol.
Through the BAC protocol, the reader and the passport agree on a key seed $xkseed$
that is then used to generate an encryption session key
as well as a MAC session key for the next protocols.
Following the description given in Figure~\ref{fig:bac}, the protocol
works as follows:
\begin{enumerate}
\item  The reader sends a constant $\getC$ to the passport that will answer by
generating a nonce, \emph{i.e.} a fresh random number.
\item  Once the reader
receives this nonce~$n_T$, it will generate its own nonce~$n_R$, as
well as a key~$k_R$ that will be used later on to derive session keys.
The reader encrypts the nonce $n_T$, its own nonce $n_R$ as well
as the key~$k_R$ with the (long-term) symmetric encryption key $\KE$. This message
$\senc(\langle n_R, \langle n_T, k_R\rangle\rangle, \KE)$ is sent to the passport
together with the associated MAC (with key $\KM$) to ensure that the encryption will be correctly
transmitted to the passport.
\item The passport performs some checks. In particular, it checks that
  the MAC has been computed using the right key $\KM$, and that the
  nonce~$n_T$ it has generated at the first step of the protocol is inside the encryption.
Once these checks have been performed, the passport sends to the
  reader a message similar to the one it has received using its own
  contribution~$k_T$.
\item Again, the reader will perform the necessary checks before
  accepting the message, and in case of success two session keys will
  be generated from the value $xkseed$ obtained by applying the
  exclusive or operator on $k_R$ and $k_T$.
\end{enumerate}

These two session keys are
used to provide confidentiality, integrity, and authentication in subsequent communications. In
particular, they are used to encrypt and MAC the messages exchanged
during the execution of the Passive and Active Authentication
protocols in order to ensure that
only parties with physical access to the passport can read the
data.
The aim of establishing fresh session keys (instead of reusing $\KE$
and $\KM$ at each session) 
is to make the passport \emph{unlinkable}, a property that will be
discussed in Section~\ref{subsec:attacks}.


\subsection{Some logical attacks on the BAC protocol}
\label{subsec:attacks}

In this section, we describe two possible attacks. These attacks are
purely logical in the sense that they do not require to break any
cryptographic primitives.

\paragraph*{Authentication issue}
First, we would like to pinpoint the fact that the order in which the
nonces $n_R$ and $n_T$ have been placed inside both encryptions 
is relevant. The careful reader will have noticed that the nonces have
been swapped: the reader encrypts $\langle n_R, \langle n_T, k_R\rangle\rangle$
whereas the passport encrypts $\langle n_T, \langle n_R, k_T\rangle\rangle$. 
The purpose of this design is to avoid a \emph{replay attack}. Indeed,
without such a swap,
a malicious user (who does not know the keys~$\KE$ and~$\KM$) 
would be able to simply replay the message he
received from the reader without decrypting it and performing the
checks. Such a message will be accepted by the reader and pass all 
checks performed by the reader. 
This means that the reader will end its
session thinking (s)he has talked with the passport identified by $\KE$
and $\KM$,
whereas this passport will not have really taken part in the protocol.
Moreover, the key seed computed at the end will be
$k_R \oplus  k_R=0$ and thus very different
  from the one that is supposed to be computed during a normal
  execution.

\paragraph*{Unlinkability issue}
Following the specification provided by the ICAO,
each
nation has implemented its own version. Unfortunately, as the specification is not
completely comprehensive, each nation’s passport has subtle
differences. In particular, the standard specifies that the passport
must reply with 
an error message to every ill formed or incorrect message from the reader, but it does not specify what the error message should be. 

For example, in the French implementation,
the passport tag replies different error messages depending on whether
the nonce in $xenc$ is not $n_T$ or $xmac$ is not a correct
MAC w.r.t. the key~$\KM$~\cite{arapinis-csf10}.
An attacker who does not know the keys $\KE$ and $\KM$ could then
trace a passport in the following way:
\begin{enumerate}
\item He eavesdrops a session between an authentic reader
  and a passport $P$ (with keys $\KE$ and $\KM$) and
stores $m = \langle xenc,  xmac \rangle$;
\item In a different session, he sends the message $m$ and waits for
  the passport's answer;
\item Then, we distinguish two cases:
\begin{enumerate}
\item if he receives a nonce error then he knows that the passport
succeeded to check the MAC and so this passport is~$P$;
\item if he receives a MAC error then he knows that the passport is
  not the one with keys $\KM$ (and $\KE$), and therefore it is not~$P$.
\end{enumerate}
\end{enumerate}
This attack makes it possible to detect when a particular passport
comes into the range of a reader, which could be, for instance, placed by a
doorway, in order to monitor when a target enters or leaves a
particular building.
To avoid the information leakage of these error messages, 
{the specification} should prescribe that, in case of failure, the
passport yields
the same
message in both situations (as it is done for instance in e-passports
from the UK).

\smallskip{}

We may note that in presence of
honest participants who follow the protocol rules, the protocol works
well, and the scenarios described above are not possible.
However, it is important to ensure that these protocols work well in
any situation, especially in the presence of 
malicious agents that may want to take advantage of the
protocol and therefore do not necessarily follow the instructions
specified by the protocols.
Verifying cryptographic protocols 
in such a hostile environment is an essential feature
which makes protocol verification a difficult task.

\section{A variety of security properties}
\label{sec:properties}

Cryptographic protocols aim at ensuring various security goals,
depending on the application. 
The two most classical  security
properties are \emph{secrecy} (also called \emph{confidentiality}) and \emph{authentication}.

\paragraph*{Secrecy}
This property concerns a message used by the protocol. This is
typically a nonce or a secret key that should not become public.
Even for this quite simple security property, several definitions
have been proposed in the literature. 
When considering the notion of (weak) secrecy, a public message 
is a message that can be learnt by the attacker.

\paragraph*{Authentication}
Many security protocols {aim at} authenticating one agent to
another: one agent should become sure of the identity of the other. 
There are also several variants of authentication. A taxonomy of these
has been proposed by Lowe in~\cite{lowe97hierarchy}.

\medskip{}

Authentication and weak secrecy are both trace properties, 
that is, statements that something bad never occurs in any execution trace of a protocol. 
Several results and tools have been developed to analyse trace properties.
However, privacy properties cannot be defined (or cannot be naturally
defined) as trace properties. They are defined relying on a notion of indistinguishability.
Intuitively, two protocols~$P$
and~$Q$ are indistinguishable if it is not possible for an
attacker to decide whether (s)he is interacting with~$P$ or~$Q$. This notion of indistinguishability 
is also used for defining a
stronger notion of secrecy, and
we may also rely on this notion of indistinguishability to compare a
protocol with an idealised version of it.
We will see in Section~\ref{subsec:equivalences} how
this notion of indistinguishability is formalised. Below, we simply
list
some security properties that can be formalised relying on such a
notion.

\paragraph*{Strong secrecy}
This notion
 is stronger than (weak) secrecy, and related to  the concept of indistinguishability.
Intuitively, strong secrecy means that an adversary cannot see any difference when the value of the secret
{changes~\cite{abadi1997secrecy,blanchet2004automatic,abadi1997calculus}}.

\paragraph*{Anonymity}
Frequently, communication between two principals reveals their
identities and  presence to third parties. Indeed, anonymity is in
general not one of the explicit goals of common authentication
protocols. However, we may want protocols that achieve this goal.
It has been informally defined in the ISO/IEC 15408-2 standard as follows:
\begin{quote}
  [Anonymity] ensures that users may use a [protocol]
without disclosing their identity.
\end{quote}
It is usually formally defined {(see~\cite{arapinis2009untraceability,schneider1996csp,chothia2006analysing})}
as the fact that an observer cannot distinguish
two scenarios where the same protocol is executed by different users.

\paragraph*{Vote privacy}
In the context of electronic voting, 
privacy means that the vote of a particular voter
is not revealed to anyone. This is one of the fundamental security
properties that an electronic voting system has to satisfy.
Vote privacy is typically defined (see \emph{e.g.}~\cite{KremerRyan2005,DKR-jcs08}) by the fact that an observer should not observe 
when two voters swap their votes, \emph{i.e.} distinguish between a situation
where 
Alice votes \emph{yes} and Bob votes \emph{no} and a situation 
where these two voters have voted the other way around.

In the context of electronic voting, some strong forms of vote-privacy
are desirable too. For instance, \emph{receipt-freeness} stipulates that 
a voter does not obtain any receipt information, which could
be used by a coercer to prove that she voted in a certain way~\cite{BenalohT94}. 
Receipt-freeness is intuitively a stronger property than privacy. 
Privacy says that an attacker cannot discern how a voter votes from
any information that the voter necessarily reveals during the course
of the election. Receipt-freeness says the same thing even 
if the voter voluntarily reveals additional information. Again,
several formal definitions of such a property have already been
proposed relying on the notion of indistinguishability (see
\emph{e.g.}~\cite{DKR-csfw06,BackesVote08}).
We may also be interested to ensure such a security property even if
the strength of the encryption has eroded
with the passage of time (which is unavoidable). 
This property, known as \emph{everlasting privacy}, has again been formalised
relying on the concept of indistinguishability~\cite{ACKR-post13}.

Note that these concepts are not specific to the electronic voting
applications and 
the definitions of privacy and receipt-freeness described above have
also been reused and adapted to model privacy and receipt-freeness in 
\emph{e.g.}
on-line auction systems~\cite{DJP-fast10,DLL-post13}.

\paragraph*{Unlinkability}
Protocols that keep the identity of their users secure may still allow an attacker
to identify particular sessions as having involved the same
principal. Such linkability attacks may, for instance, make it
possible for a third party to trace the movements of someone
carrying an RFID tag without him being able to notice anything (as the
attack on the French version of the BAC protocol described in Section~\ref{subsec:attacks}).
Intuitively, protocols are said to provide unlinkability (or
untraceability) according to the ISO/IEC 15408-2 standard, if they
\begin{quote}
  [...] ensure that a user may make multiple uses of [them]
  without others being able to link these uses together.
\end{quote}
Formally, this is often defined 
as the fact that an attacker should not be able to distinguish a scenario in which the
same agent (\ie the user) is involved in many sessions from one that
involved 
different agents in each session. Following this intuition, slightly different definitions have been proposed  (see
\emph{e.g.}~\cite{arapinis2009untraceability,arapinis-csf10,bruso2010formal,van2008untraceability}). A
comparison between these definitions may be found in~\cite{bruso2013linking}.

\bigskip{}

More generally, this notion 
allows one to express flexible notions of security by 
requiring indistinguishability between a protocol and an
idealised version of it, that magically realises the desired
properties. This is typically what is done to express security goals
of a protocol in the Universal Composability (UC)
framework. The universal composability paradigm has been quite
successful in the computational approach~\cite{CanettiFOCS01}. The idea of UC is
not, however, restricted to the computational setting, and has now 
a counterpart in symbolic models as
well~\cite{DKP-fsttcs09,UC-unruh13}.

\section{Formalising protocols and properties}
\label{sec:model}

Several symbolic models have been proposed for cryptographic
protocols. The first one  has been described by Dolev and
Yao~\cite{dolev81security} and several other models have been
proposed since then.
A unified model would enable better comparisons between the different existing
results but unfortunately such a model does not exist currently. The
reason for having several popular symbolic models probably comes
from the fact that they have to achieve two antagonistic
goals. On the one hand, models have to be as fine grained and
expressive as possible to capture a large range of applications.
One the other hand, models have to remain relatively 
simple in order to allow the design of 
verification procedures.
In order to formally define the problems we are interested in, and  to
present the existing results, we will describe one such 
model which is intuitive enough. This model is inspired 
from cryptographic calculi, and actually pretty close to the applied-pi
calculus~\cite{AbadiFournet2001}. 

\subsection{Messages}
\label{subsec:messages}

In symbolic models, messages are a key concept.
Whereas messages
are bit-strings in the real-world (and in the computational approach as
well), they are modelled using first-order terms within the symbolic model.
Atoms can be for instance nonces, keys, or agent identities. Examples of function symbols are concatenation,
asymmetric and symmetric encryptions or digital signatures.
We consider an infinite set~$\N$ of \emph{names} which are used to
represent keys and nonces (\emph{e.g.} $k$, $n$); and two infinite and disjoint sets of
\emph{variables}, denoted~$\X$ and~$\W$. Variables in~$\X$ will typically be
used to refer to unknown parts of messages expected by participants
and will be denoted $x$, $y$, $z$, \ldots
Variables in~$\W$ will be used to store messages learnt by the
attacker and will be denoted $w$, $w_1$, $w_2$, \ldots. We assume a signature~$\Sigma$, \emph{i.e.}  a set of
function symbols together with their arity. Given a signature
$\mathcal{F}$ and a set of initial data $\mathsf{A}$, we denote by
$\T(\mathcal{F}, \mathsf{A})$ the set of {\em terms} built from elements of
$\mathsf{A}$ by applying function symbols in $\mathcal{F}$. Given a
term $u$, we note $\vars(u)$ the variables that occur in $u$. A message  
is a \emph{ground} term, \emph{i.e.}  a term that does not contain any variable. 
Some works rely on a sort system for terms, and consider that only
atomic data may occur at a key position (atomic keys). 

\begin{example}
\label{ex:signature}
Consider the signature
\[
\Sigma = \{\senc,\sdec,\langle \, \rangle,
\proj_1, \proj_2, \mac, \oplus, 0\}.
\]

We use the binary symbols $\senc$ and $\sdec$ to represent symmetric encryption and decryption. Pairing is
modelled using the binary symbol $\langle \; \rangle$, whereas
projections are modelled using the unary symbols $\proj_1$ and
$\proj_2$.
The binary function symbol $\oplus$ and the constant $0$ are  used to model  the
exclusive or operator, and the binary symbol $\mac$ is used to model message authentication code.
\end{example}

There are two different ways to assign a meaning to function symbols,
which we describe next.

\paragraph{Equational theory}
To give a meaning to these function symbols,  
we associate an equational theory $\E$ to the signature $\Sigma$.
An equational theory is a $\Sigma$-congruence on terms that is closed under
substitutions of terms for variables. 
We usually require the equational theory to be closed under one-to-one
renaming (of names in $\N$), but not necessarily closed under substitutions of arbitrary
terms for names. Usually, an equational theory is generated from a
finite set of equations $M = N$ with $M, N \in \T(\Sigma,\X)$.
In this case, we have that $\E$ is closed by substitutions of terms for names.

\begin{example}
\label{ex:eq}
To reflect the algebraic properties of the exclusive or operator, as
well as the encryption/decryption and pairing/projection functions, we may
consider the equational theory generated by the following set of equations
($i \in \{1,2\}$):
\[
\begin{array}{rclcrcll}
\proj_i(\langle x_1,x_2\rangle) & =& x_i &&
                                                                  \sdec(\senc(x,y),y) &=& x\\                                                     
x \oplus 0 &=& x &\;\;\;\;\;\;\;& (x \oplus y) \oplus z &=& x \oplus (y \oplus z)&\\ 
x \oplus x &=& 0 && (x \oplus y) &=& (y \oplus x)&
\end{array}
\]
\noindent In such a case, we have that $\senc(a \oplus (b \oplus a), k) =_{\mathsf{E}} \senc(b,k)$.
\end{example}

\paragraph{Rewriting system}
Some frameworks rely on a rewriting system to give a meaning to 
function symbols. 
In such a case, function symbols in $\Sigma$ are split into
constructor/destructor symbols, namely $\Sigma = \Sigma_c \uplus \Sigma_d$,
and a rewriting system is used to reduce destructor symbols.
For instance,
assuming the signature given in Example~\ref{ex:signature}, 
let us consider $\Sigma_c = \{\senc, \langle\; \rangle, \mac, \oplus, 0\}$
and $\Sigma_d = \{\sdec, \proj_1, \proj_2\}$ together with the three
following rewriting rules:
\[
\proj_i(\langle x_1,x_2\rangle)  \to  x_i \mbox{ with $i \in
  \{1,2\}$, and    }
\sdec(\senc(x,y),y) \to x.
\]
This rewriting system allows us to rewrite terms until reaching a
message (terms that only use constructor symbols). In case such a
message cannot be reached, we say that the \emph{computation failed}.

\medskip{}

This gives us two slightly different ways to model \emph{e.g.}
symmetric encryption, with some 
fundamental differences.
Relying on a modelling using equations, an attacker will be able
to apply the decryption algorithm using a key~$k$ on top of a term
which is not a cipher-text. Considering a modelling of encryption/decryption
using a rewriting rule, 
such a computation will fail, and therefore the attacker will be able
to see whether the message is a cipher-text (encrypted with the expected
key) or not.
Some frameworks, such as the one used in the \proverif\xspace tool~\cite{ProVerif-sources},
allow both kinds of function symbols: some are equationally
defined whereas some other are defined through rewriting
rules.

For the sake of clarity, we will assume that all function symbols are
given a meaning through an equational theory only. Considering
rewriting systems would require some adaptation (for instance, the fact
that a computation fails is something that can be observed by the
attacker).

\medskip{}

{Relying on equational theories gives us enough
flexibility to model a variety of cryptographic primitives. For
instance, it is possible to model 
a \emph{blind signature} scheme (a primitive that is often used  in e-voting
 protocols) as follows~\cite{KremerRyan2005}:}
 \[
 \begin{array}{rclcrcl}
 \unblind(\blind(x,y),y) & = & x &\;\;\;& \checksign(\sign(x,y),\pk(y)) &=& x\\[1mm]
 \multicolumn{7}{c}{\unblind(\sign(\blind(x,y),z),y) = \sign(x,z)}
 \end{array}
 \]
{Intuitively, an agent can apply the function $\blind$ on a message
using a blinding factor of his choice. Then, it is possible to
retrieve the original message only for one who knows this blinding
factor. Note that the last equation also
permits one to extract a signature out of a blind signature,
but only when the blinding factor is known. }

\subsection{Assembling terms into frames}
\label{subsec:frame}

At a particular point in time, while engaging in one or more sessions
of one or more protocols, 
an attacker may know a sequence of messages (ground terms) $u_1,
\ldots, u_\ell$.
This means that he knows all messages and also their order.
So it is not enough for us to say that the attacker knows the set of
terms $\{u_1, \ldots, u_\ell\}$.
In the applied-pi calculus~\cite{AbadiFournet2001}, such a sequence of messages is
organised into a \emph{frame}, \emph{i.e.} a substitution of the form:
\[
\phi = \{w_1 \mapsto u_1, \ldots, w_\ell \mapsto
u_\ell\}.
\]
The variables $w_1,\ldots, w_\ell$ from $\W$ enable us to refer to
each message~$u_i$, and these variables will allow us to make
explicit the
order in which these messages are sent.

\smallskip{}

For modelling purposes, we split the signature $\Sigma$ into two
parts, $\Sigma_\pub$ and~$\Sigma_\priv$ (this is orthogonal to the
splitting that may have been done between constructors and destructors mentioned previously). 
An attacker builds his own messages by applying
{public function symbols (\ie in $\Sigma_\pub$)} to
ground terms he already knows and that are available through variables
from~$\W$. 
Formally, a computation done by the attacker is a \emph{recipe}, \emph{i.e.} a
term in $\T(\Sigma_\pub,\W)$. 
The application of a substitution $\sigma$ to a term $u$ is written
$u\sigma$, and we denote by $\dom(\sigma)$ its \emph{domain}.
For two recipes $R,R'$ and a frame $\phi$, we note $(R=_\E R')\phi$ when $R\phi=_\E R'\phi$.

\begin{example}
\label{ex:recipe}
Consider the signature defined in Example~\ref{ex:signature} together
with the equational theory presented in Example~\ref{ex:eq}. Let $\phi$ be the
following frame:
\[
\phi = \{w_1 \mapsto \senc(n_1,k), \;w_2 \mapsto \senc(n_2, k), \;w_3
\mapsto \senc(n_3,k), \;w_4 \mapsto k\} 
\]
We have that $R_1 = \sdec(w_1,w_4)$, $R_2 = \sdec(w_2,w_4)$, and $R_3
= \sdec(w_3,w_4)$ are three recipes. These recipes allow the attacker
to compute the messages $R_1\phi$, $R_2\phi$, and $R_3\phi$. These
terms are equal modulo $\E$ to the names $n_1$, $n_2$, and $n_3$.
\end{example}

Several notions of equivalence between processes have been introduced in the
literature to express indistinguishability, but actually they all rely
on the notion of \emph{static equivalence}.
Intuitively, an attacker can distinguish two frames
if he is able to perform a test that succeeds in one
frame, whereas it fails in the other.
More formally, we have that:

\begin{definition}
Two frames $\phi$ and $\phi'$ are in \emph{static equivalence}, written ${\phi \sim_\E \phi'}$,  when
$\dom(\phi) = \dom(\phi')$, and for any
 recipes $R_1,R_2 \in \T(\Sigma_\pub,\dom(\phi))$, we have that:\\
\null\hfill $(R_1  =_\E R_2)\phi$ if, and only if, $(R_1=_\E
R_2)\phi'$. \hfill\null
\end{definition}

\begin{example}
\label{ex:static}
Resuming Example~\ref{ex:recipe}, consider the frame:
\[
\phi' = \{w_1 \mapsto \senc(n'_1,k'), \;w_2 \mapsto \senc(n'_2, k'), \;w_3
\mapsto \senc(n'_1\oplus n'_2,k'), \;w_4 \mapsto k'\} .
\]
We have that $R \stackrel{\defi}{=} R_1 \oplus R_2$ and $R'
\stackrel{\defi}{=} R_3$ (where $R_1$, $R_2$, and $R_3$ are as defined in
Example~\ref{ex:recipe}) are two recipes such that $(R =_\E R')\phi'$
whereas this equality does not hold in $\phi$. Therefore the frames
$\phi$ and $\phi'$ 
are \emph{not} in static equivalence.

Consider the frame $\psi$ (resp.~$\psi'$) obtained by
removing  its last element $w_4 \mapsto k$
(resp. $w_4 \mapsto k'$)  from
$\phi$ (resp.~$\phi'$). We have that $\psi$ and
$\psi'$ are in static equivalence.
\end{example}

Many decidability and complexity results for 
static equivalence already exist (\emph{e.g.}~\cite{AC04Icalp,AC05,CD-jar10}),
and some of these procedures have even been implemented (\emph{e.g.}
KISS~\cite{CDK-jar10}, YAPA~\cite{BCD-tocl12}, FAST~\cite{Fast-2011}).
These results cover a wide class of cryptographic primitives 
as long as they are modelled through convergent equational theories
(\ie theories in which equations can be oriented and form a convergent rewriting system).

\smallskip{}

However, static 
equivalence is a static notion, and does not take into account the dynamic
behaviour of the underlying protocols. Static equivalence represents a
passive attacker who can
only observe messages that are sent on the public network, and is not
powerful enough to mount the attacks described in
Section~\ref{subsec:attacks}.
Even if this notion of static equivalence plays an important role for the
analysis of  security protocols in the presence of an active
attacker (\ie an attacker who may interact and interfere with the protocol), it remains
challenging to obtain decidability results for the active case, especially  in presence
of algebraic properties. The results that have already been obtained
in this direction (and the associated tools) are described in
Section~\ref{sec:methods}.

\subsection{Protocols}
\label{subsec:processes}

We assume an infinite set $\mathcal{C}h = \mathcal{C}h_0 \uplus
\mathcal{C}h^\mathsf{fresh}$ of \emph{channel names}, where
$\mathcal{C}h_0$ and  $\mathcal{C}h^\mathsf{fresh}$ are infinite and
disjoint. Intuitively, channels of $\mathcal{C}h^\mathsf{fresh}$  will
be used to instantiate channels when they are generated during the
execution of the protocol. They should not be part of a protocol specification.

\paragraph{Syntax}
Protocols are modelled through \emph{processes} built by the grammar
given below (where $c,c'\in\mathcal{C}h_0$, $x\in\X$, $n\in\N$, and, $u,u_1,u_2\in\T(\Sigma,\N\cup\X)$):
\[  
\begin{array}{lcll}
P,Q&:=&
     \phantom{|\ } 0 &\mbox{null}\\
 && |\ P \mid Q &\mbox{parallel} \\
   && |\ \In (c,x).P  &\mbox{input} \\
 &  & |\ \Out(c,u).P & \mbox{output} \\
   && |\ ! P & \mbox{replication} \\
&   & |\ \new\; n. P &\mbox{restriction}\\
   && |\ \new\; c'. \Out(c,c').P & \mbox{creation of public channel}\\
   && |\ \myIf \; u_1 = u_2 \; \myThen \; P \; \myElse \;
      Q \hspace{0.5cm} &
   \mbox{conditional}
\end{array}
\]

The process $0$ denotes the null process that does nothing.
The process ${P \mid Q}$ runs $P$ and $Q$ in parallel. 
The process $\In(c, x).P$ waits to receive a message on the public
channel $c$, 
and then continues as $P$ but with $x$ replaced by the received
message. 
The process $\Out(c,u).P$ outputs a term~$u$ on the channel $c$, 
and then continues as $P$. 
The process $!P$ executes an infinite number of copies of $P$ in
parallel.
The restriction $\new\,n.P$ is used to model the creation in a process
of new random numbers (\emph{e.g.}, nonces or key material).
The process $\new\, c'. \Out(c,c').P$ is a special construction for
creating new channels: any new channel should be made public
immediately. Intuitively, we consider here only public channels. These
fresh channel names are used to identify  a process, similarly to a
session identifier for example. 
The process \mbox{\texttt{if} {$u_1 = u_2$} \texttt{then} $P$
  \texttt{else} $Q$} runs as $P$ if the terms $u_1$ and $u_2$ are equal in the equational theory, and as $Q$ otherwise.
Note that the terms $u$, $u_1$, and $u_2$ that occur in the grammar 
may contain variables but the terms will become ground when the
evaluation will take place. Note also that $\new\, n.P$ and
$\In(c,x).P$ are binding constructs, respectively for the name~$n$ and
for the variable~$x$, and in both cases the scope of the binding is~$P$.

We consider only a fragment of the applied-pi calculus. In particular, we
  do not allow channel passing nor internal communication. Such a
  calculus will be (almost) sufficient to present all the existing results. 
A
  \emph{protocol} is a ground process, \emph{i.e.} a process whose variables
  are in the scope of an input. 

\smallskip{}

For the sake of clarity, we often omit the null
process, and we omit the $\myElse$ branch of a conditional when it
contains the~$0$ process.

\begin{example}
\label{ex:bac-proc}
We are now able to model the French variant of the tag's role of the
BAC protocol (see Figure~\ref{fig:bac}) as a process
parameterised by two (long-term) keys~$ke$, and~$km$.
\[
\begin{array}{rcl}
P_\mathrm{Tag}(ke,km)\;  &:=& \In(c_T,z).  
\new~n_T.
\Out(c_T, n_T).\In(c_T, x).\\
&&\myIf\;\mathsf{mac}(\pi_1(x),km)=\pi_2(x)\\
&&\myThen\; \myIf\;\pi_1(\pi_2(\sdec(\pi_1(x),ke)))=n_T\;\\
&&\phantom{then} \myThen\; \new~k_T.\Out(c_T,\langle m,\mathsf{mac}(m,km)\rangle)\\
&&\phantom{then}\myElse \; \Out(\mathsf{error_{Nonce}}) \\
&&\myElse \;\Out(\mathsf{{error}_{Mac}}) \\
\end{array}
\]
where $m=\senc(\langle n_T, \langle \pi_1(\sdec(\pi_1(x),ke)),k_T
\rangle\rangle,ke)$.
If $P_\mathrm{Reader}(ke,km)$ is the process modelling the reader, the protocol BAC
with many readers and tags that can play arbitrary many sessions can
be modelled through the following process:
\[
P_{\mathrm{BAC}}=\,!\;\new~ke.\new~km.~!\;(P_\mathrm{Tag}(ke,km)\;|\;
P_{\mathrm{Reader}}(ke,km)).
\]
\end{example}

Configurations represent processes having
already evolved by \emph{e.g.}  disclosing some terms to the
environment.
\begin{definition}
  A \emph{configuration} is a pair $(\p; \phi)$ where
$\p$ is a multiset of \emph{ground processes}; and
$\phi = \{w_1 \mapsto u_1, \ldots, w_n \mapsto u_n\}$ is a
    \emph{frame}.
\end{definition}

We implicitly assume that null processes are removed from a configuration.
The applied-pi calculus as introduced in~\cite{AbadiFournet2001} does not introduce
this notion of configuration but considers instead the notion of
\emph{extended processes} together with a notion of structural
equivalence to identify processes that are identical up to some
rearrangement of their structure (\emph{e.g.} $P \mid Q$ and $Q \mid P$).
A configuration can be seen as a more canonical way to represent
an extended process, and this will avoid us to rely on a notion of
structural equivalence.

A configuration is said to be \emph{initial} when it 
does not use channel names from $\mathcal{C}h^\mathsf{fresh}$.

In some cases, it is helpful to know which process has executed a
given observable action.
Hence, some methods and tools we will discuss in Section~\ref{sec:methods} consider the class
of the {\em simple processes}.

\begin{definition}
\label{def:simple-proc}
  A {\em simple process} is a process of the form:
  \[
!\new~c_1.\Out(c'_1,c_1).B_{1} ~|~\cdots~|~ !\new~c_n.\Out(c'_{n},c_n).B_{n}
  \mid   B_{n+1}~|~\cdots~|~B_{n+k}
  \]
  where all $c_i, c'_i$ are distinct and for any $1\le i\le n+k$,
  $B_i$ is a \emph{basic process} built on channel $c_i$, where
  a {\em basic process} on channel~$c$ is a process built using the
  following  grammar:
\[
B, B' := 0 \mid \In(c,x).B \mid \Out(c,u).B \mid \new \,n.B \mid \myIf
\; u_1 = u_2 \; \myThen \; P \; \myElse \; Q
\]
\end{definition}

This class is reasonable given that the attacker often knows with whom he is communicating. For simple processes, this is 
reflected by the fact that concurrent processes use different
channels that are observable by the attacker.

\smallskip{}

\paragraph{Semantics}
The semantics is given by a labelled transition system (LTS) on
configurations (see
Figure~\ref{fig:semantics}). 
This labelled operational semantics allows one to avoid the
quantification over all contexts when analysing a protocol in presence
of an arbitrary attacker.

The rules are quite standard and correspond to the intuitive meaning of the
syntax given in the previous section. When a process emits a message,
we distinguish two cases.
The rule {\sc Out} corresponds to the output of a term by some
process: the corresponding term is added to the frame of the current
configuration, which means that the attacker can now access the sent
term. The rule {\sc Ch} corresponds to the special case of an output
of a freshly 
generated channel name. In such a case, the channel is not added to
the frame but 
it is implicitly assumed known to the attacker, as all the channel names.

These rules define the relation $\lrstep{\;\ell\;}$, where~$\ell$ is either
an input, an output, or a silent action~$\tau$. The
relation~$\lrstep{\; \tr \;}$ where $\tr$ denotes a sequence of labels
is defined in the usual way, whereas the relation $\LRstep{\; \tr' \;}$
on configurations is defined by: $K \,\LRstep{\;\tr' \;}\, K'$ if, and only
if, 
either $K = K'$ and $\tr' = \epsilon$ (the empty trace); or
there exists a sequence $\tr$ such that $K \lrstep{\; \tr \;} K'$
and~$\tr'$ is obtained by erasing all occurrences of the silent
action~$\tau$ in $\tr$. 

\smallskip{}

Given an initial configuration $K_0 = \config{\p}{\phi}$, we define its set of
traces as follows:
\[
\trace(K_0) = \{(\tr,\phi') ~|~ K_0
\,\LRstep{\;\tr\;}\, \config{\p'}{\phi'} \mbox{ for some
  configuration $\config{\p'}{\phi'}$}\}.
\]

 \begin{figure*}[ht]
\[
\begin{array}{rcl}
\multicolumn{3}{l}{\mbox{\sc Then}}\\
\config{\{\mbox{\texttt{if} $u_1 = u_2$ \texttt{then} $P_1$
     \texttt{else} $P_2$}\}\uplus\p}{\phi}
  &\lrstep{\tau} & \config{P_1\uplus\p}{\phi}  \;\;\;\; \hfill \mbox{when $u_1 =_\E u_2$}\\[1mm]
\multicolumn{3}{l}{\mbox{\sc Else}}\\
   \config{\{\mbox{\texttt{if} $u_1 = u_2$ \texttt{then} $P_1$
     \texttt{else} $P_2$}\}\uplus\p}{\phi}
   & \lrstep{\tau} & \config{P_2\uplus\p}{\phi} \hfill\mbox{when $u_1 \neq_{\E} u_2$}\\
 \end{array}
\]
\vspace{-0.2cm}
\[
\begin{array}{l}
\mbox{\sc In}\\
  \config{\{\In(c,z).P\}\uplus\p}{\phi}
   \lrstep{\In(c,R)} 
  \config{P\{z \mapsto R\phi\} \uplus\p}{\phi}  \hfill \mbox{where $\vars(R) \subseteq \dom(\phi)$}
  \\[2mm]
\mbox{\sc Out}\\
  \config{\{\Out(c,u).P\}\uplus\p}{\phi}
   \lrstep{\Out(c,w)} 
                           \config{P\uplus\p}{\phi\cup\{w\mapsto u\}} \;\;\;
                           \hfill \mbox{where $w \in \W$ is fresh}\\[2mm]
\mbox{\sc Ch}\\
  \config{\{\new\; c'. \Out(c,c').P\} \uplus \p}{\phi}  \lrstep{\mathsf{outCh}(c,ch)}
  \config{P\{c'\mapsto ch\} \uplus \p }{\phi}\;\;\;\\
\hfill \mbox{where $ch\in\mathcal{C}h^\mathsf{fresh}$ is fresh}
\end{array}
\]
\vspace{-5pt}
\[
\begin{array}{ll}
 \mbox{\sc New} &\config{\{\new\; n.P\} \uplus \p}{\phi} \lrstep{\tau} 
\config{P\{n \mapsto n'\}\uplus \p }{\phi} \;\;\;\;\;\mbox{where $n' \in
  \N$ is fresh}
  \\[1mm]
\mbox{\sc Par} &\config{\{P_1 \mid P_2\} \uplus \p}{\phi} \lrstep{\tau} 
  \config{\{P_1,P_2\} \uplus \p}{\phi} \\[1mm]
\mbox{\sc Repl}\;\;&
  \config{\{!P\} \uplus \p}{\phi}  \lrstep{\tau}
  \config{\{!P, P\} \uplus \p }{\phi} \\[1mm]
\end{array}
\]
\caption{Semantics}
\label{fig:semantics}
\end{figure*}

\begin{example}
\label{ex:bac-exe}
 Continuing Example~\ref{ex:bac-proc}, we consider the following
 configuration
\[
K_A=(\{P_\mathrm{Tag}(ke^A,km^A)\};\{w_0\mapsto
\langle m_0,\mac(m_0,km^A)\rangle\})
\]
 where
 $m_0=\senc(\langle n_R^0,\langle n_T^0,k_R^0\rangle\rangle, ke^A)$.
 This configuration represents a scenario where the attacker has eavesdropped
 a message $\langle m_0, \mac(m_0,km^A)\rangle$
 from a previous session between a reader and  Alice's tag with keys 
 $ke^A$, and~$km^A$; and now the attacker is again in presence of
 Alice's tag. We have that
   $K_A\LRstep{\tr_A}
   (\emptyset;\phi_A)$ where 
\begin{itemize}
\item $\tr_A
   =\In(c_T,\getC).\Out(c_T,w_1).\In(c_T,w_0).\Out(c_T,w_2)$;
   and
\item $\phi_A = \{ w_0\mapsto
\langle m_0,\mac(m_0,km^A)\rangle;\; w_1 \mapsto \langle
m_A,\mac(m_A,km^A)\rangle; \; w_2 \mapsto
\mathtt{error}_{\mathsf{Nonce}}\}$ (for some $m_A$).
 \end{itemize}
\end{example}

{Our calculus has similarities with the spi calculus~\cite{AbadiGordon99}. 
The key difference concerns the way in which cryptographic primitives
are handled. 
The spi calculus has a fixed set of built-in primitives (namely,
symmetric and public key encryption), while our calculus allows a wide
variety of primitives 
to be defined by means of an equational theory as in the applied-pi calculus.
}
Process algebras are not the only way to model protocols. We may at
least mention the  multiset rewriting (MSR) model that has been introduced in~\cite{CervesatoDLMS99} to model
reachability properties (\emph{e.g.} weak secrecy, authentication), and the
strand space model~\cite{ThayerHG99-jcs} that comes with an appealing graphical representation,
and some proof techniques. These two models have been recently extended
to capture an equivalence based property similar to
the notion of diff-equivalence that we will introduce in the following
section.

\subsection{Equivalences}
\label{subsec:equivalences}

In order to express the security properties introduced in
Section~\ref{sec:properties}, we need to formally define the notion of
indistinguishability we are interested in.  Intuitively, 
two processes are indistinguishable if an attacker has no way to tell them apart.
A natural starting point is  to say 
that processes~$P$ and~$Q$ are indistinguishable 
if they can output on the same channels, no matter the context in which they are placed. 
The quantification over contexts makes this definition hard to use in
practice. Therefore indistinguishability notions, which are
more suitable for
both
manual and automatic reasoning, have been proposed. All these notions rely
on a labelled transition semantics as the one presented in
Section~\ref{subsec:processes} to  reason about protocols that may
interact with an environment that models an arbitrary attacker.

Here, we make the choice to directly present a labelled
  semantics together with equivalence notions that are based on this
  semantics and therefore avoid the quantification over contexts
  required when using a reduction semantics. 
Actually linking these two semantics and their associated
  notions of equivalence is not an easy task. 
Starting with the pioneering work of Milner and Sangiorgi~\cite{MiSa92},
this problem has been
  addressed for different calculi and different notions of equivalence
  in several papers
(\emph{e.g.} pi-calculus, spi-calculus~\cite{AbadiG98,BorealeNicolaPuglieseJoC02},
applied-pi calculus~\cite{AbadiFournet2001}, and
psi-calculus~\cite{psi-calculus}).

\medskip{}

\paragraph*{Trace equivalence}
The notion that seems to be the most appropriate to capture the notion
of indistinguishability we are interested in is the notion of trace
equivalence. 

\begin{definition}
  \label{def:trace-equiv}
  Let $K_P$ and $K_Q$ be two initial configurations, ${K_P \sqsubseteq_t K_Q}$ if
  for every $(\tr,\phi) \in \trace(K_P)$, there exists $(\tr', \phi')
  \in \trace(K_Q)$ such that ${\tr = \tr'}$ and~${\phi \sim
    \phi'}$.
  We say that~$K_P$ and~$K_Q$ are \emph{trace equivalent}, denoted by $K_P
  \approx_t K_Q$, if ${K_P \sqsubseteq_t K_Q}$ and~${K_Q \sqsubseteq_t K_P}$.
\end{definition}

\begin{example}
\label{ex:bac-equ}
In order to formalise whether the attacker is able to distinguish
between Alice's tag and Bob's tag,
one may want to check if $K_A$ is trace equivalent to
\[K_B=(\{P_\mathrm{Tag}(ke^B,km^B)\};\{w_0\mapsto \langle
m_0,\mac(m,km^A)\rangle\})\]
This equivalence actually fails to hold.
We have that $(\tr_B,\phi_B)\in\trace(K_B)$ for some trace~$\tr_B$
that has exactly the same observable actions as $\tr_A$. However, the
only possible
resulting frame is $\phi_B$ (for some message $m_B$):
\[
\{ w_0\mapsto
\langle m_0,\mac(m_0,km^A)\rangle;\; w_1 \mapsto \langle
m_B,\mac(m_B,km^B)\rangle; \; w_2 \mapsto
\mathtt{error}_{\mathsf{Mac}}\}
\]
It is easy to see that $\phi_A \sim_\E\phi_B$ does \emph{not} hold.
Indeed, using recipes $R_1=w_2$ and
$R_2=\mathtt{error}_{\mathsf{Nonce}}$, we have  that 
$(R_1=_\E R_2)\phi_A$ but 
$(R_1\neq_\E R_2)\phi_B$.

Hence, just by looking at the second output of the tag and checking whether it is equal
to the public constant $\mathtt{error}_{\mathsf{Nonce}}$, the attacker is able to
learn if he was interacting with Bob or Alice.
This formalises the unlinkability attack discussed in Section~\ref{subsec:protocol} for the specific
case of two sessions.

Note that, in case the two error messages  $\mathtt{error}_{\mathsf{Nonce}}$ and
$\mathtt{error}_{\mathsf{Mac}}$ were equal as in the UK version, one would have $K_A^\mathsf{UK}\approx_t K_B^\mathsf{UK}$.
This is a non-trivial equivalence that can be established using
\emph{e.g} the \APTE\xspace tool presented
in Section~\ref{sec:methods}.
\end{example}

\medskip{}

\paragraph*{Labelled bisimilarity}
Showing  trace equivalence properties is a very difficult
task. The notion of labelled bisimilarity for the spi-calculus has been introduced to
approximate trace equivalence~\cite{AbadiGordon99}.
The fact that labelled bisimilarity  is based on a notion of
step-by-step simulation between processes makes this notion sometimes
easier to establish directly.

\begin{definition}
\label{def:labelled}
\emph{Labelled bisimilarity} $\approx$ is the largest symmetric relation $\mathcal{R}$
on configurations such that $\config{\p}{\phi} \mathrel{\mathcal{R}}
\config{\q}{\psi}$ implies:
\begin{enumerate}
\item static equivalence: $\phi \sim \psi$;
\item if $\config{\p}{\phi} \lrstep{\tau} K_P$, then
  $\config{\q}{\psi} \LRstep{\tau} K_Q$ and
  $K_P\mathrel{\mathcal{R}} K_Q$ for some~$K_Q$;
\item  if $\config{\p}{\phi} \lrstep{\alpha} K_P$, then
  $\config{\q}{\psi} \LRstep{\alpha} K_Q$ and
  $K_P\mathrel{\mathcal{R}} K_Q$ for some $K_Q$.
\end{enumerate}
Two initial configurations $K_P$ and $K_Q$ are labelled bisimilar if
$K_P \approx K_Q$.
\end{definition}

It is well-known that labelled bisimilarity implies trace equivalence
whereas the converse is false in general. Actually, it has been proved
in~\cite{TCS2013-Vincent} that these two notions coincide for a large
class of processes that include in particular the class of simple
processes as described in Definition~\ref{def:simple-proc}.

\medskip{}

\paragraph*{Diff-equivalence}
Another notion of equivalence that has been extensively used in the
context of cryptographic protocols verification is the notion of \emph{diff-equivalence}.
Such a notion is defined on bi-processes that are pairs of processes
that have the same structure and differ only in the choice of terms
they use.
The syntax is similar to the one introduced in
Section~\ref{subsec:processes} but each term $u$ has to be replaced by
a bi-term written $\choice[u_1,u_2]$ (using \proverif\xspace syntax).
Given a bi-process $P$, the process $\fst(P)$ is obtained by replacing
all occurrences of $\choice[u_1,u_2]$ with $u_1$. Similarly, $\snd(P)$
is obtained by replacing $\choice[u_1,u_2]$ with $u_2$. These notations
are also used for bi-frames.

The semantics of bi-processes is defined as expected via a relation
that expresses when 
and how a bi-configuration may evolve. A bi-process reduces if, and only if,
both sides of the bi-process 
reduce in the same way: \emph{e.g.}
a conditional has to be evaluated in the same way on both sides.
For instance, the {\sc Then} and {\sc Else} rules are as follows:
\[
\begin{array}{l}
{\mbox{\sc Then}}\\
\config{\{\mbox{\texttt{if} $\choice[u_1,u_2] = \choice[v_1,v_2]$ \texttt{then} $Q_1$
     \texttt{else} $Q_2$}\}\uplus\p}{\phi}
  \lrstep{\tau}_{\mathsf{bi}}  \config{Q_1\uplus\p}{\phi} \\
\hfill \mbox{when $u_1 =_\E v_1$ and $u_2 =_\E v_2$}\\[1mm]
{\mbox{\sc Else}}\\
   \config{\{\mbox{\texttt{if} $\choice[u_1,u_2] = \choice[v_1,v_2]$ \texttt{then} $Q_1$
     \texttt{else} $Q_2$}\}\uplus\p}{\phi}
    \lrstep{\tau}_{\mathsf{bi}}  \config{Q_2\uplus\p}{\phi} \\
\hfill\mbox{when $u_1 \neq_{\E} v_1$ and $u_2 \neq_{\E} v_2$}\\
 \end{array}
\]

When the two sides of the bi-process reduce in different ways, the bi-process
blocks. The relation $\LRstep{\tr}_{\mathsf{bi}}$ on bi-processes is
therefore defined as for processes. This leads
us to the following notion of diff-equivalence.

\begin{definition}
\label{def:diff-equiv}
An initial bi-configuration $K_0$ satisfies \emph{diff-equivalence} if for
every bi-configuration
$K = \config{\p}{\phi}$ such that $K_0 \,\LRstep{\tr}_{\mathsf{bi}}\, K$
for some trace $\tr$, we have that:
\begin{itemize}
\item $\fst(\phi) \sim \snd(\phi)$; 
\item if $\fst(K) \lrstep{\alpha} A_L$ then there exists a
  bi-configuration~$K'$ such that $K \lrstep{\alpha}_{\mathsf{bi}} K'$ and
  $\fst(K') =A_L$ (and similarly for $\snd$).
\end{itemize} 
\end{definition}

As expected, this notion of diff-equivalence  is actually stronger than
the usual notion of labelled bisimilarity, and thus trace equivalence.
It may be the case that the 
two sides of the bi-process reduce in different
ways (e.g. taking two different branches in a conditional) but still
produce the same observable actions.
This strong notion of diff-equivalence happens to be sufficient to establish some
interesting equivalence-based properties such as strong secrecy, and
anonymity. However, this notion is actually too strong to establish
for example vote privacy for many interesting e-voting protocols~\cite{DKR-jcs08},
or unlinkability as defined in~\cite{arapinis-csf10}.

For instance, looking back to Example~\ref{ex:bac-equ} (when
error messages are equal), it can be shown that
$K_A^\mathsf{UK}\approx_t K_B^\mathsf{UK}$. On the other hand, 
$K_A^\mathsf{UK}$ and
$K_B^\mathsf{UK}$ are not related by diff-equivalence. Indeed, the
first three observable actions of $\tr_A / \tr_B$ 
are executable, but this results in a bi-process with a conditional that
evaluates differently on both sides. Therefore, even
if the error message outputted on both sides is the same,
diff-equivalence does not hold.

\section{Methods and tools for verifying equivalence-based properties}
\label{sec:methods}
Modelling protocols using the symbolic approach allows one to benefit from
machine support through the use of various existing techniques, ranging from 
model-checking to resolution and rewriting
techniques.
{Aiming at machine support is really relevant since manual proofs are error-prone, tedious and
hardly verifiable. Moreover, new protocols are developed quite frequently and need to be verified quickly.}
Nevertheless, verifying a security property in such a setting
(and especially those  expressed using the notion of equivalence) remains a
difficult problem which is 
actually undecidable~\cite{huttel2003deciding,CCD-tocl15}.

\subsection{Bounded number of sessions}
\label{subsec:bounded}

In order to design decision procedures,
a reasonable assumption is to bound the number of protocol sessions
{(\ie forbid replication)},
thereby limiting the length of execution traces. Under such an
assumption,  the first decision procedure towards automatic 
verification of equivalence between protocols dates back
to~\cite{huttel2003deciding}, where a fragment of
the spi calculus (no replication, no else branch) is considered.
Note that, even under this assumption, infinitely many traces remain, since each input may be fed infinitely
many different messages. This issue has been tackled in various ways using
forms of symbolic execution and  the development of dedicated
procedures.
{Obtaining a symbolic semantics to avoid 
potentially infinite branching of execution trees due to inputs from
the environment is often a first step towards automation of
equivalence. Depending on the 
expressivity of the calculus and the way
its semantics is given,  this task can be quite cumbersome (\emph{e.g.} applied-pi
calculus~\cite{DKR-jcs09}, spi calculus~\cite{DSV03,Borgstrom09}, psi calculus~\cite{BorgstromGRV15}) and
sometimes only leads to
 incomplete procedures.}

A table summarising the main features of existing tools dedicated to
bounded verification 
is given in Table~\ref{fig:tools-bounded}.

\subsubsection{Constraint solving approaches}

Baudet targets the decision of security of protocols against off-line guessing attacks defined using 
static equivalence between open frames (\ie frames with some unknown
parts constrained with some deducibility and equality constraints)~\cite{baudet2005deciding}.
The main novelty of his work was to design a constraint solving procedure
that is not only able to solve satisfiability problems (sufficient for
reachability properties) but also to establish equivalences
(\ie two systems have the same sets of solutions), which  are needed when one wants to verify equivalence-based security properties.
This is done for a user-defined equational theory given in the form of a
subterm convergent rewriting system (\ie convergent and such
  that the
  right-hand side of each rewriting rule is actually a syntactic
  subterm of the left-hand side). 
As a result, this work allows for verifying trace equivalence of
simple processes (with no else branch) for all the standard primitives~\cite{TCS2013-Vincent}.

A shorter proof of the result by Baudet is
given in~\cite{chevalier10}. It is shown that if two processes are not equivalent, then
there must exist a small witness of non-equivalence, and a decision
procedure can be derived by checking every possible small witness. 
The main issue with all the results mentioned
so far is practicality. Consequently,
they have not been implemented.

\medskip{}

A decision procedure for a stronger notion of trace equivalence (namely {\em open bisimulation})
has been proposed in~\cite{tiu2010automating} and implemented in the
tool \SPEC\footnote{http://www.ntu.edu.sg/home/atiu/spec/index.html}.
The procedure deals with a fixed set of cryptographic primitives,
namely symmetric encryption and pairs, and protocols with no
else
branch.
The procedure is sound and complete w.r.t. open
bisimulation (a notion that is strictly stronger than trace equivalence~\cite{tiu2007trace})
and its termination is proved.
The attacker's deductive ability is modelled as logical rules in sequent calculus,
and procedures deciding message deduction and  message indistinguishability are defined
as proof-search strategies. 
Finally, the proposed procedure iteratively builds an open bisimulation
from the two initial processes by symbolically executing them
and checks that possible instantiations are coherent on both sides.

\medskip{}

For a fixed but richer set of cryptographic primitives
(\ie symmetric/asymmetric encryptions, signature, pair, and hash functions),
a different procedure, presented in~\cite{CCD-ccs11} (improved version of~\cite{CCD-ijcar10}), allows to decide equivalence of two
{\em sets} of constraint systems that may also feature {\em disequality tests}. 
Dealing with disequality tests and sets of constraint systems 
is needed in the presence of
protocols with else branches
 (different symbolic executions may be associated to a single
symbolic trace). {Actually, the procedure presented in~\cite{CCD-ccs11}
allows for slightly more general processes than those presented in
Section~\ref{sec:model} since it deals with private channels and internal
communications.}
The tool \APTE~\cite{Cheval-tacas14} implements the procedure
described in~\cite{CCD-ccs11}.
This procedure explores all possible symbolic
traces and computes all possible resulting symbolic constraint systems on both sides.
This forward symbolic exploration of two processes is finite since all
symbolic traces have a bounded length
and the exploration is finitely branching since inputs are abstracted away by variables and constraints.
The procedure then checks the symbolic equivalence of all the
resulting pairs of
sets of constraint systems.
Recently, this procedure has been further extended to deal with some
forms of {\em side-channel} attacks regarding the
length of messages~\cite{CCP-cav13},
and the computation time~\cite{cheval2015timing}.

\begin{table}[t]
\begin{center}
\begin{tabular}{|>{\centering}p{2cm}|>{\centering}p{3cm}|>{\centering}p{3cm}|>{\centering}p{3cm}|}
\hline
& \APTE~\cite{Cheval-tacas14}  & \AKISS~\cite{chadha2012automated} & \SPEC~\cite{tiu2010automating}  
 \tabularnewline\hline\hline
\hline
 Equivalence     & $\approx_t$ & $\approx_{\mathit{ct}}$,
                                     $\approx_{\mathit{ft}}$ &  open bisim. \tabularnewline[2mm] \hline
\vskip1pt Primitives      & \vskip1pt standard &  convergent with finite variant
                                  &  pair \& sym. encryption 
                                                          \tabularnewline[2mm]\hline
Class of protocols & \vskip2ptfull & linear role with equality
                                            tests & 
                                                    linear role with
                                                    filtering   \tabularnewline[2mm]
  \hline
 Input syntax   & \multicolumn{2}{c|}{\mbox{applied-pi calculus}}
  &\vskip-4pt spi calculus   \tabularnewline \hline
 \vskip6pt  Termination       & \vskip6pt  proved & proved for sub. convergent &
                                                                      \vskip6pt proved
                                                                       \tabularnewline[2mm]
                                                                                 \hline
 Exploration       &   \multicolumn{3}{c|}{forward}   \tabularnewline[1mm]
    \hline\hline
\end{tabular}
\end{center}
  \caption{Main features of existing tools (for a bounded number of sessions)}
  \label{fig:tools-bounded}
\end{table}

\subsubsection{Resolution-based approaches}

The procedure described in~\cite{chadha2012automated} deals 
with rich user-defined term algebras provided that they can be defined using a convergent rewriting system 
enjoying the {\em finite variant
  property}~\cite{comon2005finite}. This property
basically requires that it is possible to finitely pre-compute
possible normal forms of terms with variables.
This especially includes all subterm convergent equational theories. 
In the setting of~\cite{chadha2012automated}, protocols are modelled as sets of symbolic traces
with equality tests. 
Further, the authors of~\cite{chadha2012automated} use first-order Horn clauses to model all possible
instantiations of symbolic traces,
and they rely on a saturation procedure to put all clauses into {\em solved forms}.
Finally, this finite description of all possible concrete executions
is used to decide equivalence between the two processes under study.
This procedure is actually able to check an over-approximation (called
$\approx_{ct}$) and an under approximation (called $\approx_{ft}$)
of trace equivalence, and it has been shown that $\approx_{ct}$ actually coincides
with trace equivalence for a large class of processes (the class of
determinate processes) that typically includes simple processes.
This procedure has been implemented in the tool \AKISS\footnote{http://akiss.gforge.inria.fr} and has been
effectively tested on several examples including checking vote privacy
of an electronic voting protocol relying on the blind signature
primitive.
Recently,  termination of the procedure has been established for
subterm convergent theories~\cite{CCCK16}.

\medskip{}

Systems we are interested in are highly concurrent and existing
methods and tools 
naively explore all possible symbolic interleavings
causing the so called state-explosion problem. This problem seriously limits the practical impact of those tools.
Recent works~\cite{BDH-concur15,baelde2014reduced}
have partially addressed this issue by developing dedicated partial order reduction techniques to dramatically
reduce the number of interleavings to explore. They have been implemented in \APTE~and brought significant speed-up.
Actually, they are generic enough to be applicable to any method
as long as it performs forward symbolic executions.


\subsection{Unbounded number of sessions}

The decidability results mentioned in the previous section 
analyse equivalence for a bounded number of sessions only, that is
assuming that protocols are executed a limited number of times. 
This is of course a strong limitation. Even if no flaw is found when a
protocol is executed $n$ times, 
there is absolutely no guarantee that the protocol remains secure when
it is executed $n + 1$ times. Therefore, despite the difficulty of the
problem in the general case, several solutions have been proposed for an
unbounded number of sessions.
A table summarising the main features of existing tools dealing with an unbounded number of sessions
is given in Table~\ref{fig:tools-unbounded}.

\subsubsection{Decidability results}
\label{subsubsec:decidability}
It is well-known that replication (allowing us to encode an unbounded
number of sessions) very quickly leads to undecidability even when
considering
the simple and well-known weak
secrecy property.
Therefore, obtaining decidability results can only be achieved under
various restrictions.

One of the first decidability results  for checking trace 
equivalence of protocols for an unbounded number
of sessions is due to Chrétien \emph{et al.}~\cite{icalp2013,CCD-tocl15}.
They consider a limited fragment of protocols, namely
\emph{ping-pong protocols}, with standard primitives but
pairs, and at most one variable per protocol rules. 
Even if the secrecy preservation problem is known to
be decidable in this setting~\cite{rta03}, it turns out that checking equivalence is
undecidable.
Then, considering determinate protocols, they establish decidability
through a characterisation of equivalence of
protocols 
in terms of equality of languages of (generalised, real-time)
deterministic pushdown automata. Note that this result only holds for 
a restricted signature, and
names can only be used to produce randomised cipher-texts.
Very recently, the algorithm for checking equivalence of
deterministic pushdown automata has been implemented~\cite{Geraud-outil}, and the translation from protocols to
pushdown automata has been implemented too, yielding the first
prototype able to decide trace equivalence
considering an unbounded number of sessions~\cite{CCD-tocl15}.

\medskip{}

 Assuming finitely many nonces and keys, another decidability result
 has been obtained in~\cite{CCD-concur14}. The primitives that are considered are
 pairs, and symmetric encryption only, but they go beyond ping-pong
 protocols and consider the class of
 simple protocols. In order to derive a strong typing
 result that drastically limits the shape and size of messages needed
 to mount an attack, the authors introduce the notion of \emph{type-compliance} for
 a protocol. 
This notion generalises the idea of tagging
 as introduced by Blanchet in~\cite{blanchetTag03}, and avoids
 ambiguity in the interpretation of the origin of any message sent on
 the network. From this typing result, they derive a 
decision procedure for trace equivalence for an unbounded number of
sessions for type-compliant protocols without nonces.

\medskip{}

Actually, the typing result mentioned above has also been used to
derive the first decidability result for trace equivalence in presence
of unlimited fresh nonces~\cite{CCD-csf15}. 
Such a decidability result inherits the conditions introduced above
(type-compliance, restricted signature),
and develops in
addition a
notion of \emph{dependency graph}. This notion formally abstracts the dependencies
between the actions occurring in a protocol
specification. Then, considering acyclic protocols (\ie those for which the
dependency graph is acyclic) which is intuitively related to protocols
without loops in their well-typed executions, 
decidability of trace equivalence  is established.
These procedures have not been implemented yet.


\subsubsection{Procedures for checking diff-equivalence}
\label{subsubsec:diff-equiv}

As said before, the problem of checking trace equivalence for rich class of protocols is undecidable.
To circumvent this undecidability result, 
many works aim
at developing 
procedures (not necessarily completely automatic) that are sound
w.r.t. trace equivalence but not complete. 
Moreover, termination is not guaranteed. 
The main idea is to merge the protocols under study into a so-called
bi-process, and to consider a strong form of equivalence, namely
\emph{diff-equivalence} as described in Definition~\ref{def:diff-equiv}.
This method has first been presented
in~\cite{BlanchetAbadiFournetLICS05} and implemented in the
\proverif\xspace
tool. Recently, this technique has been
integrated into the verification tools \tamarin~\cite{basin2015automated}  and \maudenpa~\cite{santiago2014formal} that have
been extended to deal with equivalence properties. 
The main limitation of all these results is the fact that the tools
are not able to analyse trace equivalence (but only diff-equivalence).
Thus, these tools are not well-suited in general to analyse several
privacy-related properties such as (strong)
unlinkability~\cite{arapinis-csf10}, and vote privacy~\cite{DKR-jcs08}.

\medskip{}


The method presented in~\cite{BlanchetAbadiFournetJLAP07} and
implemented in \proverif\footnote{http://prosecco.gforge.inria.fr/personal/bblanche/proverif/} represents bi-processes that are given in
input using Horn clauses
(performing some well-chosen approximations, and thus losing completeness).
Then, a dedicated resolution algorithm tries to resolve those Horn
clauses. Cryptographic primitives 
are decomposed into: reduction rules
and a union of linear equational theories (\ie each equation has
the same variables on both sides)
and convergent theories (\ie terminating and confluent). This
formalism is flexible enough to model for instance different flavours
of encryptions (symmetric, asymmetric, randomised, \ldots),
signature, and blind signature,
but excludes exclusive-or, and more generally associative and
commutative operators.
The resulting tool is quite efficient, and terminates on many
examples.

As already mentioned, diff-equivalence is strictly stronger than
trace equivalence. Basically, the two processes
have to be executed exactly in the same way,
notably for internal rules, 
whereas
 the attacker cannot observe such details. 
This problem has been partially tackled in~\cite{ChevalBlanchetPOST13}
by pushing away the evaluation of conditionals
into terms. Nevertheless, the problem remains in general (\eg for
interleavings of conditionals and observable actions). 

To extend the class of equivalences and protocols 
that can be automatically verified
by \proverif, several extensions have been proposed.
For instance,
ProSwapper\footnote{http://www.bensmyth.com/proswapper.php}) has been designed to consider
cryptographic protocols that require barrier synchronisation  (also
called phases) to
achieve security objectives~\cite{DRS-ifiptm08}. The ProSwapper extension allows the
algorithm to go beyond diff-equivalence by rearranging 
bi-processes. This extension has been shown  particularly useful to establish
vote privacy for several electronic-voting protocols. 
More recently, theoretical foundations have been provided for this technique
and soundness of this extension has been proved~\cite{2016-verifying-observational-equivalence}.
{A reduction result to get rid of some particular equations
  (that cannot be handled
by the \proverif\xspace tool) has been devised in~\cite{ArapinisBR12}. Relying on it, 
a first automated proof of  privacy for the protocol Pr\^et \`a Voter
(that uses re-encryption and associative/commutative operators)
has been carried out with success.}

\medskip{}

\begin{table}[t]
\begin{center}
\begin{tabular}{|>{\centering}p{2cm}|>{\centering}p{2.6cm}|>{\centering}p{3.3cm}|>{\centering}p{3.3cm}|}
\hline
& \proverif~\cite{BlanchetAbadiFournetJLAP07}
  &\maudenpa~\cite{santiago2014formal} &
                                         \tamarin~\cite{meier2013tamarin} \\
 \tabularnewline
\hline\hline
\hline
 Equivalence     &  \multicolumn{3}{c|}{diff-equivalence} \tabularnewline[2mm] \hline
\vskip4pt Primitives      & \vskip1pt linear + convergent  &
                                                             convergent
                                                             with
                                                             finite
                                                             variant\\
                                                             (inc. XOR, A.G.)
                                  &   convergent with finite variant\\
                                    (inc. DH)
                                                          \tabularnewline[2mm]\hline
Class of protocols &\vskip2pt full & linear role with filtering & \vskip2pt
                                                  full + state \tabularnewline[2mm]
  \hline
 Input syntax   & applied-pi calculus & \vskip0.5pt strand spaces &
                                                                  \vskip0.5pt
                                                                    multiset
                                                                    rewriting   \tabularnewline \hline
Termination       & \multicolumn{3}{c|}{may diverge}
                                                                       \tabularnewline[2mm]
                                                                                 \hline
 Exploration       &  resolution & \multicolumn{2}{c|}{backward}   \tabularnewline[1mm]
    \hline\hline
\end{tabular}
\end{center}
  \caption{Main features of existing tools (for an  unbounded number of sessions)}
  \label{fig:tools-unbounded}
\end{table}

Recently, the approach behind the \tamarin\xspace verification tool~\cite{tamarin} has been extended to deal with
equivalence-based properties. In this approach,
protocols are modelled as
multiset rewriting (MSR) systems. This allows one to model a
rich class of protocols that may feature
else branches and allow the storage of some data from one session to another. The framework
supports a rich term algebra including subterm convergent theories, and
Diffie-Hellman exponentiation.  The proposed algorithm exploits the finite
variant property~\cite{comon2005finite} to get rid of some equations,
and it builds
on ideas from strands spaces and proof normal forms. It basically
performs a backward  search from attacks states. \tamarin\xspace provides
two ways of constructing proofs: an efficient, fully automated mode
that uses heuristics to guide proof search, and an interactive mode. 
The interactive mode enables the user to explore the proof states
using a graphical interface. 
The \tamarin\xspace tool has been used to analyse different security
properties on many protocols.
However,  regarding equivalence, the
tool is less mature and has only been used on a few examples; the main one being a
stateful TPM protocol (namely the TPM envelope protocol) on which a
strong secrecy property has been established.

\medskip{{}

The \maudenpa\xspace tool has also been recently extended~\cite{santiago2014formal} to deal with
bi-processes (called {\em synchronous product}) and diff-equivalence.
Their semi-decision procedure is able to deal with a very large class of term algebras
(as soon as they have the {\em finite variant property} as defined in~\cite{escobar2010folding})
like Abelian groups, exclusive-or, and exponentiation.
However, it can only be applied to linear role scripts  with filtering
over inputs (and therefore
does not handle protocols with else branches).
Regarding equivalence, only a few case studies have been
performed. 
The approach of~\cite{santiago2014formal} suffers from termination problems, especially when
considering primitives such as exclusive-or. Regarding equivalence,
their main example is a proof of absence of guessing attacks on a version of the EKE protocol (that
relies on standard primitives only).

\subsubsection{Some other results}
In many cases, existing methods and tools are not sufficient to carry out fully
automated proofs. 
On the other side, fully manual proofs are tedious,  error-prone, and hardly verifiable.
For instance, previous
works
gave a manual and formal proof of vote privacy 
for Helios~\cite{cortier2011attacking} and the Norwegian e-voting protocol~\cite{cortier2012formal}. Those proofs are not automated mainly because existing tools
are not able to deal with  an unbounded number of voters and complex equational theories 
featuring for instance homomorphic encryption.
For such proofs, one has to exhibit complex bisimulation relations and show static equivalence
of infinite families of frames.

Actually, it is also possible to combine manual and automatic proofs.
For instance, in~\cite{arapinis2014privacy}, the authors establish an
unlinkability property on a fixed version of the TMSI reallocation procedure used in mobile
telephony systems. As pointed out in the paper, no tool could, at this time, deal
with both stateful protocols and an equivalence-based property like
unlinkability. 
Thus, they exhibit a manually-built bisimulation and discharge static
equivalence verification to \proverif.

This idea has also  been applied to analyse some privacy properties
(namely unlinkability and
forward privacy) on a very restricted class of stateful RFID
protocols~\cite{bruso2010formal}. In this work, 
single-step protocols that only use hash functions as cryptographic
primitives are considered.
In such a restricted setting, the authors introduce the
notion of frame independence which is closely related to the notion of
static equivalence between frames. Then, they provide 
conditions under which unlinkability and forward privacy hold. They
perform several case studies and establish those conditions (up to
some arbitrary bound) using \proverif.
More recently, a new method~\cite{HBD-sp16} based on sufficient conditions for unlinkability and anonymity
allows to automatically verify such security properties for unbounded number of sessions for protocols that were
out of the scope of existing tools (\eg BAC  protocol and some RFID protocols).
Instead of improving the tools and their precisions, such approaches rather focus on security properties of interest
and devise sufficient conditions that are checkable much more easily.

 \section{Conclusion}
\label{sec:conclu}

The results obtained so far are not completely satisfactory.
The protocols that
are deployed nowadays in various applications rely on operators and
primitives that can still not be handled by existing verification algorithms
and tools. For instance, electronic voting protocols often
rely on complex primitives in order to achieve their goals
(\emph{e.g.} homomorphic encryption~\cite{cortier2011attacking,cortier2012formal}), and RFID protocols that have
power-consumption constraints often use some low-level operators with
algebraic properties (\emph{e.g.} exclusive or
operator~\cite{van2009algebraic}). Despite some advances that
have been done in this direction (see \emph{e.g.}~\cite{DKP-ijcar12}), 
analysing these protocols is still out of scope
of the existing algorithms and tools.

Due to the complexity of the verification problem (especially when
considering equivalence-based properties), a promising approach seems
to devise methods (with tool support) that are sound but not
necessarily complete.
However, we advocate 
verification techniques that go beyond the notion of
diff-equivalence. Learning from previous experience, we deem it
acceptable to provide tools 
with some hints, in order to guide them in their attempt to
establish
the equivalence property under study.
The main
difficulty is probably to find a reasonable way to allow interactions between the
user and the tool during the verification process.





\paragraph*{Acknowledgements}
This work has been partially supported by the ANR project Sequoia ANR-14-CE28-0030-01.

\section{References}
\bibliographystyle{elsarticle-num}
\bibliography{refs}

\end{document}